\documentclass[twocolumn, 
notitlepage,
reprint,
amsmath,amssymb,
aps,
prl,
]{revtex4-1}

\usepackage{amsmath}
\usepackage{amssymb}
\usepackage{graphicx}
\usepackage{dcolumn}
\usepackage{bm}
\usepackage{color}

\DeclareMathOperator{\Tr}{Tr}

\DeclareMathOperator{\diag}{diag}

\DeclareMathOperator{\Imag}{Im}

\newcommand{\av}{\textbf{a}}

\newcommand{\J}{\textbf{J}}

\newcommand{\rv}{\textbf{r}}

\begin{document}
	\title{Photonic refrigeration from time-modulated thermal emission}
	
	\author{Siddharth Buddhiraju}
		\author{Wei Li}
	\author{Shanhui Fan}%
	\email{sbuddhi@stanford.edu, shanhui@stanford.edu}
	\affiliation{%
		Ginzton Laboratory, Department of Electrical Engineering, Stanford University, Stanford, CA
	}%
	\date{\today}
	
	\begin{abstract}
 Active photonic cooling is of significant importance to realize robust, compact, vibration-free, all-solid-state refrigeration. Currently proposed photonic cooling approaches are based on luminescence and impose stringent requirements on luminescence efficiency. We propose a new photonic cooling mechanism arising from temporal modulation of thermal emission. We show that this mechanism has a high thermodynamic performance that can approach the Carnot limit and yet does not rely on luminescence. Further, our work opens exciting new avenues in active, time-modulated control of thermal emission for cooling and energy harvesting applications.
\end{abstract}

	\maketitle

Active photonic cooling is of significant importance with the potential to realize robust, compact, vibration-free, all-solid-state refrigeration. To date, the proposed photonic cooling approaches, including laser cooling \cite{epstein1995observation,seletskiy2010laser,zhang2013laser} and electroluminescent cooling \cite{tauc1957share,chen2015heat,xiao2018electroluminescent,zhu2019near}, are all based on luminescence and impose stringent requirements on luminescence efficiency. Recent advances in nanophotonic control of thermal emission \cite{lenert2014nanophotonic,raman2014passive,ilic2016tailoring,greffet2002coherent,liu2011taming,khandekar2015radiative,li2018nanophotonic} have offered a promising approach to manipulate photon emission without the luminescence process. In this Letter, we propose a photonic refrigeration technique based on time-modulation of thermal emission. This technique results in a mechanism of refrigeration with a significantly higher performance than laser cooling of solids, while also overcoming the stringent quantum efficiency requirements of electroluminescent cooling. Our work opens exciting new avenues in active, time-modulated control of thermal emission for active cooling and energy harvesting applications.\\

Laser cooling and electroluminescent cooling are the two main existing approaches to active photonic refrigeration. The mechanism of laser cooling is based on an anti-Stokes luminescence up-conversion process \cite{pringsheim1929zwei}, where the energy difference between a luminescence photon and a pump photon is supplied by the lattice phonon, resulting in cooling. Net laser cooling has been demonstrated in rare earth metal doped crystals \cite{epstein1995observation,seletskiy2010laser} and more recently in II-VI semiconductors \cite{zhang2013laser}. However, the performance of laser cooling is inherently limited to be several orders of magnitude below the Carnot bound due to the small energy difference between the luminescence photon and the pump photon \cite{seletskiy2016laser}. Alternatively, electroluminescent cooling has been suggested to realize photonic cooling \cite{tauc1957share,berdahl1985radiant,chen2015heat,chen2016near,zhu2019near} with the potential to overcome such limitations on performance. In positive electroluminescent cooling, a positive chemical potential for photons enables thermal emission at a radiant temperature that is much higher than the thermodynamic temperature. However, no positive electroluminescent cooling has been demonstrated to date due to the stringent requirements on the luminescence efficiency. More recently, negative electroluminescent cooling was proposed \cite{chen2016near} and demonstrated \cite{zhu2019near}, but it suffers from the limitation of low power density.\\

As another photon emission process, thermal radiation results from a direct conversion of heat into photon emission due to the thermally induced fluctuations of particles or quasi-particles. Particularly, recent advances in using sub-wavelength nanophotonic structures to control fundamental properties of thermal radiation has offered a tremendous number of opportunities \cite{li2018nanophotonic,greffet2002coherent,liu2011taming,khandekar2015radiative} and enabled new applications such as passive radiative cooling \cite{raman2014passive}. However, so far, almost all of the existing work on thermal radiation control has focused on static systems, which can only perform passive cooling \cite{raman2014passive}, where heat flows from a high-temperature to a low-temperature object. \\

\begin{figure}
    \centering
    \includegraphics[scale=0.38]{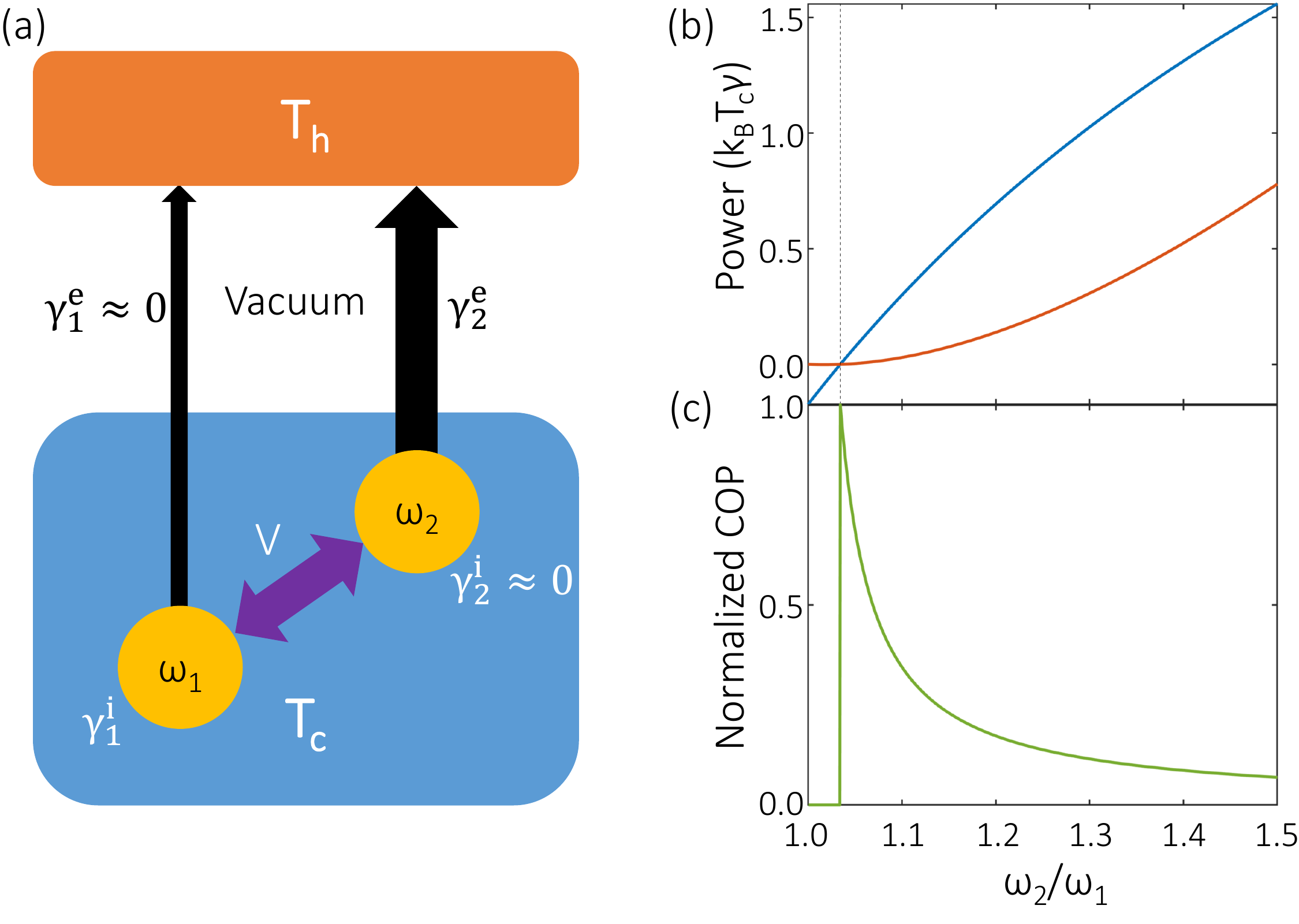}
    \caption{(a) General setup of thermal photonic refrigerator operating between a cold side at $T_c$ and a hot side $T_h$. The cold side comprises modes 1 and 2 at frequencies $\omega_{1,2}$. Mode 1 has an absorption rate $\gamma_1^i$ and a small radiative coupling rate $\gamma_1^e$ to the hot side. Mode 2 has a small absorption rate $\gamma_2^i$ and a large radiative coupling rate $\gamma_2^e$ to the hot side, as indicated by the thick arrow. Thermally generated photons populate mode 1, which are up-converted by time modulation to mode 2, indicated by the purple arrow, and then emitted out. Similarly, photons received in mode 2 are down-converted to mode 1 and absorbed. For illustration purposes, in this Figure and in the discussion of coupled-mode theory, we assume $\gamma_1^e=\gamma_2^i=0$ and $\gamma_1^i=\gamma_2^e=\gamma$. (b) Net cooling power (blue curve) and work input (red curve) normalized to $k_BT_c\gamma$, as a function of the ratio of the frequencies of the two modes for $T_h=300$ K and $T_c=290$ K at $V=2\gamma$. For $\omega_2/\omega_1> T_h/T_c$, a net outward flux from the cold side is observed, indicating cooling. (c) Coefficient of performance (COP) of the refrigerator normalized to the Carnot limit of $T_c/(T_h-T_c)$. At the onset of cooling, the COP of the refrigerator achieves the Carnot limit.}
    \label{fig:intro}
\end{figure}

In this paper we consider thermal emission from temporally modulated systems. In recent years, temporal modulation of the refractive index has offered exciting opportunities to manipulate photons. Such modulation, via the conversion of photon frequencies, can achieve optical isolation and circulation \cite{zongfu2009,kejie2012,sounas2017} and the breaking of symmetry between emission and absorption \cite{hadad2016}. While temporally modulated systems such as electro-optic modulators have been widely used in optical information processing and communication systems, the thermodynamic implications of temporal modulation have not been previously explored. In this work, we develop a statistical temporal coupled mode theory to show that temporal modulation of a thermal photonic system achieves refrigeration. Further, by a rigorous fluctuational electrodynamics approach, we verify the predictions of our theory and numerically demonstrate photonic refrigeration by computing the heat transfer from a temporally modulated body maintained at a certain temperature to a passive thermal emitter at a higher temperature.\\

Consider a cavity with modes 1 and 2 at frequencies $\omega_{1,2}$, respectively, with $\omega_1<\omega_2$, as shown schematically in Fig. \ref{fig:intro}(a). The amplitudes in the two modes $a_{1,2}$ are normalized such that $|a_{1,2}|^2$ represent the energy in the modes. The modes have internal loss rates $\gamma_{1,2}^i$ due to absorption and are in thermal equilibrium with a heat bath at temperature $T_c$. The modes also radiatively couple to an external heat bath at temperature $T_h (\ge T_c)$ via coupling rates $\gamma_{1,2}^e$. By the fluctuation-dissipation theorem, there are associated compensating noise sources \cite{otey2010thermal,zhu2013temporal} $n_{1,2}^i$ for internal loss and $n_{1,2}^e$ for external radiative coupling, respectively. The strength of these noise sources are defined by $\langle n_{1,2}^{i*}(\omega) n^i_{1,2}(\omega')\rangle = 2\pi\delta(\omega-\omega')\Theta(\omega,T_c)$ and $\langle n_{1,2}^{e*}(\omega) n^e_{1,2}(\omega')\rangle = 2\pi\delta(\omega-\omega')\Theta(\omega,T_h)$, where $\langle\cdot\rangle$ denotes a thermal ensemble average and $\Theta(\omega,T)=\hbar\omega/(\exp(\hbar\omega/k_BT)-1)$ is the Planck distribution at frequency $\omega$ and temperature $T$. Defining $\av = (a_1,a_2)^T$, $\textbf{n}_i = (n_1^i,n_2^i)^T$ and $\textbf{n}_e=(n_1^e,n_2^e)^T$, the time evolution of the two modes is described by
\begin{align}
-i\frac{d}{dt}\av &= \left(H_0-i\Gamma_i-i\Gamma_e + M(t)\right)\av + D_i\textbf{n}_i + D_e\textbf{n}_e.
\label{eq:evol}
\end{align}
Here, $H_0=\diag(\omega_1,\omega_2)$ is the Hamiltonian of the unmodulated, closed system. $D_i = \diag(\sqrt{2\gamma_1^i},\sqrt{2\gamma_2^i})$, $D_e = \diag(\sqrt{2\gamma_1^e},\sqrt{2\gamma_2^e})$, $\Gamma_{i}=D_{i} D_{i}^\dag/2$ and $\Gamma_e = D_{e} D_{e}^\dag/2$. The operator $M(t)$ describes the modulation-induced coupling between the modes. Here, we assume that the cavity is modulated by an index modulation proportional to $\cos(\Omega t)$, where $\Omega=\omega_2-\omega_1$. Then, in the simplest approximation that includes the rotating-wave approximation, $M(t)$ is given by (see Supplementary Information (SI), Section I)
\begin{equation}
    M(t) = \begin{pmatrix}
    0 & \sqrt{\frac{\omega_1}{\omega_2}}Ve^{-i\Omega t} \\ \sqrt{\frac{\omega_2}{\omega_1}}V^*e^{i\Omega t} & 0
    \end{pmatrix}, \label{eq:modulation}
\end{equation}
where $V$ is related to the strength of the index modulation. We emphasize that $M(t)$ is $\emph{not}$ Hermitian. In our formalism, the modal amplitudes $a_{1,2}(t)$ are normalized with respect to energy, and such time modulation preserves the total number of photons \cite{zongfu2009, kejie2012} and not the total energy. \\

We now show that the system shown in Fig. \ref{fig:intro}(a), which is described by  Eqs. \eqref{eq:evol}-\eqref{eq:modulation}, can achieve photonic refrigeration. For illustration purposes, we first consider the simplest case: in the setup of Fig. \ref{fig:intro}(a), we assume that mode 1 has no radiative coupling to the high temperature heat bath, i.e., $\gamma_1^e=0$. Mode 2 is assumed to have a nonzero external radiative coupling rate, but has no internal loss, i.e., $\gamma_2^i=0$. Therefore, the thermal emission and absorption of the unmodulated system in this ideal limit is zero. For simplicity, we take the remaining rates to be equal, i.e., $\gamma_1^i=\gamma_2^e=\gamma$. For this ideal system, the flux of the thermal emission from the cold side is (see SI Sec. IV, based on Secs. II-III)
\begin{equation}
    P_{out} = 4\gamma^2|V|^2\frac{\omega_2}{\omega_1}\int d\omega\frac{1}{|Z_2(\omega)|^2}\Theta(\omega-\Omega,T_c), \label{eq:pem}
\end{equation}
while the flux received from the hot side at $T_h$ is 
\begin{equation}
    P_{in} = 4\gamma^2|V|^2\int d\omega \frac{1}{|Z_1(\omega)|^2}\Theta(\omega+\Omega,T_h), \label{eq:pabs}
\end{equation}
for a work input of 
\begin{equation}
    \dot{W} = 4\gamma^2|V|^2 \int d\omega \left[\frac{\Omega}{\omega_1}\frac{\Theta(\omega,T_c)}{|Z_1(\omega)|^2} - \frac{\Omega}{\omega_2}\frac{\Theta(\omega,T_h)}{|Z_2(\omega)|^2}\right]
    \label{eq:cmtWork}
\end{equation}
where $Z_{1,2}(\omega)= (\omega-\omega_{1,2}-i\gamma_1)(\omega-\omega_{1,2}-i\gamma_2)-|V|^2$. As seen from Eqs. \eqref{eq:pem}-\eqref{eq:cmtWork}, when the modulation is turned on, i.e., $V\neq 0$, a fraction of the thermally generated photons from mode 1 are up-converted to mode 2 and emitted. These photons carry power $P_{out}$ away from the low-temperature reservoir and therefore their emission constitutes a cooling mechanism.  Similarly, a fraction of the photons received by mode 2 are down-converted to mode 1, where they are absorbed. These photons carry power $P_{in}$ into the low-temperature heat bath and thus their absorption constitutes a heating mechanism. In Fig. \ref{fig:intro}(b), we plot the net cooling given by $P_{out}-P_{in}-\dot{W}$ (blue curve) and the work input $\dot{W}$ (red curve) as a function of the ratio $\omega_2/\omega_1$ for $V=2\gamma$. We note that net cooling starts to occur when 
\begin{equation}
    \frac{\omega_2}{\omega_1} \ge \frac{T_h}{T_c}. \label{eq:cond1}
\end{equation}
As $\omega_2/\omega_1$ increases beyond the threshold value of $T_h/T_c$, the cooling power also increases. In Fig. \ref{fig:intro}(c), we plot the coefficient of performance (COP), defined as the ratio between the cooling power and the work input, as a function of $\omega_2/\omega_1$. We observe that the COP reaches the Carnot bound of $T_c/(T_h-T_c)$ at the threshold condition of Eq. \eqref{eq:cond1}, and decreases as $\omega_2/\omega_1$ increases beyond the threshold. \\ 

The threshold condition for $\omega_2/\omega_1$ in Eq. \eqref{eq:cond1} can also be derived analytically from Eqs.\eqref{eq:pem}-\eqref{eq:cmtWork} (SI Section IV). Here, we provide an intuitive argument. For simplicity, we assume the classical limit of $k_BT_{c,h} \gg \ \hbar\omega_{1,2}$. In the unmodulated cavity, the number of thermal photons in mode 1 is $k_BT_c/\hbar\omega_1$, while the number of thermal photons in mode 2 is $k_BT_h/\hbar\omega_2$, due to its radiative coupling to the high-temperature heat bath and the lack of internal loss in mode 2. When modulation is turned on, since the rate of up- and down-conversion for an individual photon is equal \cite{zongfu2009,kejie2012}, net cooling will be observed when $k_BT_c/\hbar\omega_1 \ge k_BT_h/\hbar\omega_2$, which leads to the threshold condition of Eq. \eqref{eq:cond1}. When the condition of Eq. \eqref{eq:cond1} is met, for each photon that is emitted by the modulated system, the system at $T_c$ experiences cooling by $\hbar\omega_1$. On the other hand, the work input per emitted photon is the energy difference of the two modes, $\hbar\omega_2-\hbar\omega_1$. Therefore, the COP of such a refrigerator is given by
\begin{equation}
COP  = \frac{\omega_1}{\omega_2-\omega_1} \le \frac{T_c}{T_h-T_c}, \label{eq:carnot}
\end{equation}
where the inequality follows from Eq. \eqref{eq:cond1}. This upper bound indicates that modulation-induced refrigeration obeys the Carnot limit on performance. Interestingly, the value of COP for this ideal refrigerator is independent of the modulation strength $V$. A rigorous derivation of the Carnot bound on the COP is included in the SI (Section V). \\ 

\begin{figure*}
    \centering
    \includegraphics[scale=0.52]{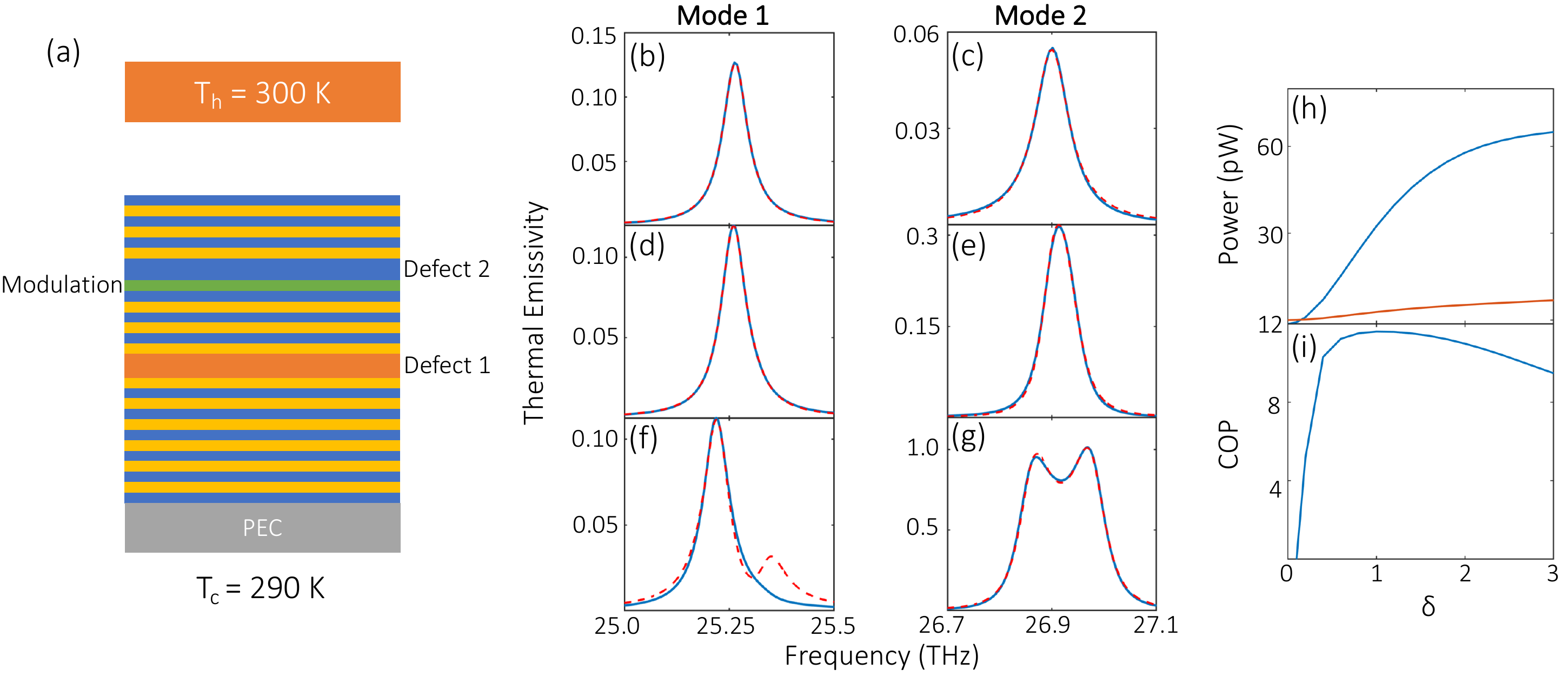}
    \caption{(a) Physical structure to demonstrate active cooling as induced by modulation. The cold side is maintained at a temperature $T_c=290$ K. The structure consists of a Bragg mirror with $\epsilon_1=14$ (blue) and $\epsilon_2=4$ (yellow) layers, each 1 $\mu$m thick. In the Bragg mirror, two defects are introduced with thicknesses 1.8 $\mu$m (defect 1) and 1.6 $\mu$m (defect 2). The material composing defect 1 is an effective medium comprising silicon carbide and the high index layer ($\epsilon_1$) in a 1:9 ratio, with parameters given in the main text. Defect 2 has a permittivity of $\epsilon_1=14$. The dielectric constant of a low index layer in between the two defects (green layer) is modulated as $\epsilon(t) = 4 + \delta\cos(\Omega t)$. The structure is bounded by a perfect electric conductor (PEC) on the bottom, and faces a hot side at $T_h=300$ K comprising the same medium as Defect 1. (b)-(c) Emission spectrum of the two modes in the unmodulated structure (solid blue lines) superimposed by a fit using coupled mode theory (red dotted lines). (d)-(e) Emission spectrum of the two modes for $\delta =0.4$ and $\Omega = 2\pi\cdot 1.64$ THz, superimposed by a coupled mode theory fit (red dotted lines), demonstrating a very good agreement. (f)-(g) emission and coupled-mode theory fit for a larger $\delta=2.0$ and the same modulation frequency. (h) Net cooling (blue) of the cold side and work input (red) to the modulated layer for the single channel, as a function of the strength of the modulation strength $\delta$. (i) the COP corresponding to the cooling shown in (h).}
    \label{fig:matching}
\end{figure*}

Motivated by the results of our coupled-mode theory, we proceed to consider a physical structure whose radiative thermal properties can be tailored by temporal modulation of its dielectric function, shown in Fig. \ref{fig:matching}(a). The structure consists of a one-dimensional photonic crystal constructed using two materials with dielectric constants $\epsilon_1=14$ (blue layers) and $\epsilon_2 = 4$ (yellow layers). Such a 1D photonic crystal with a large index contrast possesses a bandgap for waves that can propagate in vacuum. To introduce two modes in the bandgap of the photonic crystal, we create two defect layers with thicknesses greater than those of the remaining layers, indicated by `Defect 1' and `Defect 2' in Fig. \ref{fig:matching}(a). The material constituting the Defect 1 (orange layer) is assumed to be a narrowband absorber. Such narrowband absorbers help suppress parasitic heating arising from frequencies away from the modulated modes under consideration. As an example, assume Defect 1 comprises a medium that is a random mixing of silicon carbide and a lossless high-index medium of $\epsilon_1=14$ in a 1:9 ratio. Using the Maxwell-Garnett approximation, the dielectric constant of such a medium is then $\epsilon(\omega)=0.1\times\epsilon_\infty(\omega_{LO}^2-\omega^2-i\omega\gamma)/(\omega_{TO}^2-\omega^2-i\omega\gamma) + 0.9\times 14$, where $\epsilon_\infty=6.7$, $\omega_{LO}=1.83\times 10^{14}$ rad/s, $\omega_{TO}=1.49\times 10^{14}$ rad/s and $\gamma=8.97\times 10^{11}$ rad/s. Further, a layer in between the defects, marked in green, experiences a temporal modulation given by $\epsilon(t) =  \epsilon_2 + \delta\cos(\Omega t)$, where $\delta$ is the modulation strength and $\Omega$ is the modulation frequency. This structure is maintained at $T_c = 290$ K and faces a narrowband emitter in the far-field, composed of the same material as Defect 1 and at a temperature of $T_h=300$ K. \\

To perform calculations of thermal emission and absorption, we extend the formalism of radiative heat transfer \cite{polder1971theory,kaifeng2018} to include time-varying dielectric functions. This formalism combines rigorous coupled wave analysis \cite{whittaker1999scattering, inampudi2019} with the fluctuation-dissipation theorem \cite{henry1996quantum} to compute thermal emission for spatio-temporally modulated layered structures. In our system, the modulated layer does not have loss and hence by itself does not generate thermal radiation. For the lossy layer, the fluctuation-dissipation theorem has the form \cite{kaifeng2018}
\begin{align}
\langle J_\alpha(\rv, \omega)J_\beta^*(\rv', &\omega')\rangle = \frac{4}{\pi}\omega\Theta(\omega,T)\times \nonumber \\
&\frac{\epsilon_{\alpha\beta}(\rv,\omega)-\epsilon_{\beta\alpha}^*(\rv,\omega)}{2i}\delta(\rv-\rv')\delta(\omega-\omega')
\end{align}
where $\alpha,\beta=\{x,y,z\}$, $\J(\rv,\omega)$ is the current source at position $\rv$ and frequency $\omega$ that produces thermal fluctuations, and $\epsilon(\rv,\omega)$ is the dielectric tensor of the structure. Within this formalism, the net heat transfer between two bodies at temperatures $T_h$ and $T_c$ separated by a vacuum gap is given by
\begin{align}
    \Delta P = \int_0^\infty d\omega \int_{-\infty}^\infty \frac{dk_x}{2\pi} \int_{-\infty}^\infty \frac{dk_y}{2\pi} \Big[\Phi_f(\omega,k_x,k_y)\times \nonumber \\ 
    \Theta(\omega,T_c) - \Phi_b(\omega,-k_x,-k_y)\Theta(\omega,T_h)\Big], \label{eq:emittedpower}
\end{align}
where $(k_x, k_y)$ are the wavevector components parallel to the layers and $\Phi_{f,b}(\omega,k_x,k_y)$ are the Poynting flux spectra in the vacuum gap generated by sources in the cold and hot sides, respectively. For passive reciprocal structures, $\Phi_f(\omega,k_x,k_y) = \Phi_b(\omega,-k_x,-k_y)$. In this system, $\Phi_f(\omega,k_x,k_y) \neq \Phi_b(\omega,-k_x,-k_y)$ due to the presence of an actively modulated region.  In addition to the flux of thermally generated photons, we compute the work done by the modulation directly from Maxwell's equations, given by (see SI, Section VI)
\begin{align}
\dot{W} = \int_0^\infty d\omega \frac{2}{\pi}\omega\Theta(\omega,T)\int d\rv \int d\rv' \Imag\Tr\big[\mathcal{W}\hat{\epsilon}(\rv)\times \nonumber \\
\mathcal{G}(\rv,\rv')\delta_{\omega,\omega'}\Imag\left[\epsilon(\rv')\right]\mathcal{G}^\dag(\rv,\rv')\big], \label{eq:wexact}
\end{align}
where $\mathcal{W}$, $\delta_{\omega,\omega'}$ and $\hat{\epsilon}$ are matrices defined by $\mathcal{W}_{nm} = (\omega + m\Omega)\delta_{nm}$, $\delta_{\omega,\omega'} = \delta_{n=0,m=0}$ and $\hat{\epsilon}_{nm}(\rv) = \frac{\Omega}{2\pi}\int_0^{2\pi/\Omega}\epsilon(\rv, t)e^{-i(n-m)\Omega t} dt$, with $\hat{\epsilon}^\dag=\hat{\epsilon}$ in the modulation layer. $\mathcal{G}(\rv,\rv')$ is the Green's function for the electric field at $\rv$ in the modulated layer originating from a source at $\rv'$ in the lossy layers. The operator $\delta_{\omega,\omega'}$ ensures that thermal photons are generated only at $\omega$ but not at the sideband frequencies, since the lossy layer at $\rv'$ is unmodulated. We also note that the expression for work in Eq. \eqref{eq:cmtWork} is a coupled-mode theory version of the general formula given by Eq. \eqref{eq:wexact}.\\

As a first numerical demonstration, we fit our coupled-mode theory to direct numerical calculations of thermal emission into vacuum, for the structure shown in Fig. \ref{fig:matching}(a) without the hot side. In Fig. \ref{fig:matching}(b)-(c), we plot in blue the emissivity of the two modes in the unmodulated structure in the $(k_x,k_y)=(0,0)$ channel. We extract the parameters $\omega_{1,2}$, $\gamma_{1,2}^i$ and $\gamma_{1,2}^e$ by fitting the emissivity profiles, shown in red dotted lines (parameter values listed in SI, Section VII). In this structure, the lossy Defect 1 layer is further away from the top surface as compared to the lossless Defect 2 layer. Thus, $\gamma_1^e$ and $\gamma_2^i$ are much smaller than the other two rates, and therefore the thermal emission of the unmodulated system is very low. We then introduce a modulation of $\delta\cos(\Omega t)$ in the green layer in Fig. \ref{fig:matching}(a), where $\delta=0.4$ and $\Omega=\omega_2-\omega_1=2\pi\cdot 1.64$ THz. In Fig. \ref{fig:matching}(d)-(e), we plot the emissivities of the two modes from the numerical calculation in blue lines and fit them using our coupled-mode theory in red dotted lines, exhibiting a very good agreement. With modulation, the emissivity near $\omega_1$ is suppressed, whereas the emissivity near $\omega_2$ is enhanced as compared with the unmodulated system. In fact, for a larger modulation of $\delta=2.0$, shown in Fig. \ref{fig:matching}(f)-(g), the emission near $\omega_2$ becomes super-Planckian: the emissivity, which is defined as the emitted power density normalized against a blackbody at the same temperature, begins to exceed unity. The results here demonstrate that there is significant up-conversion induced by the temporal modulation. In addition, we observe modulation-induced Rabi splitting \cite{jerry2018} of the modes for $\delta=2.0$, resulting in dips in thermal emission near the frequencies where emission was maximum in the unmodulated system. We note that the emissivity under any modulation strength can be numerically computed accurately. On the other hand, coupled-mode theory, which is a first-order perturbation theory, is not a physically accurate model of the underlying dynamics for large modulation strengths and a coupled-mode theory fit should only be used to guide intuition. The parameters of the coupled mode theory fit for Figs. \ref{fig:matching}(b)-(g) are provided in the SI (Section VII).\\

To demonstrate cooling for this single channel, in the presence of the narrowband emitter on the hot side, in Fig. \ref{fig:matching}(f), we plot the net cooling of the cold side (blue curve) and the work input to the modulated region (red curve) as a function of the modulation strength $\delta$. In Fig. \ref{fig:matching}(g), we plot the corresponding COP. It is seen that the system of Fig. \ref{fig:matching}(a) does achieve cooling for the single channel under consideration with a large COP, reaching a maximum value of about 11. For reference, the Carnot limit on performance for the temperatures used in our setup is $T_c/(T_h-T_c)=29$, although this limit is attained only at net zero cooling power. \\

\begin{figure}
    \centering
    \includegraphics[scale=0.55]{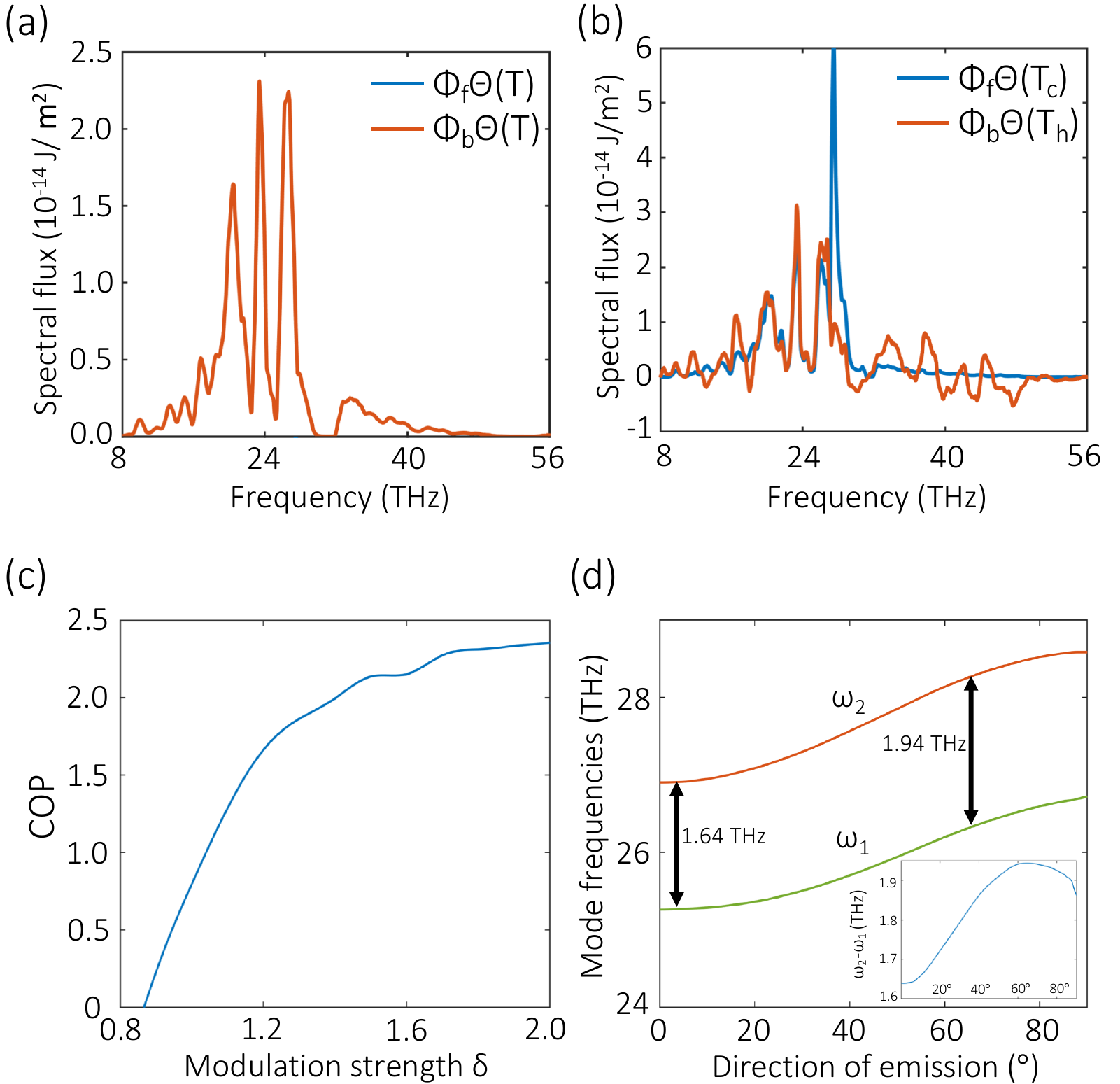}
    \caption{(a) Spectral heat flux in both forward (blue) and reverse (red) directions in the unmodulated structure of Fig. \ref{fig:matching}(a) overlay perfectly on top of each other when the two sides are at the same temperature of 300 K, indicating that $\Phi_f=\Phi_b$. (c) Spectral heat flux in the structure modulated with $\delta=1.5$, $T_c=290$ K and $T_h=300$ K, showing a significant difference between $\Phi_f$ and $\Phi_b$ and a strong enhancement of $\Phi_f$ due to photon up-conversion. (c) Thermodynamic COP of the modulated structure as a function of modulation strength $\delta$. (d) Variation in the resonant frequencies of the two resonant modes in the structure of Fig. \ref{fig:matching}(a) as the emission direction varies from normal ($0^\circ$) to glancing angle ($90^\circ$). The inset schematically depicts the frequency separation $\omega_2-\omega_1$ as a function of the emission angle.}
    \label{fig:fullsystem}
\end{figure}

Now, we proceed to demonstrate that the system of Fig. \ref{fig:matching}(a) exhibits refrigeration even after integration over all propagating channels $(k_x,k_y)$ in Eq. \eqref{eq:emittedpower} and all frequencies. Defining
\begin{equation}
    \Phi_{f,b}(\omega) = \int_{-\infty}^\infty \frac{dk_x}{2\pi}\int_{-\infty}^\infty \frac{dk_y}{2\pi} \Phi_{f,b}(\omega,k_x,k_y),
\end{equation}
in Fig. \ref{fig:fullsystem}(a), we plot the spectral heat flux $\Phi_f(\omega)\Theta(\omega,T)$ (blue curve) and $\Phi_b(\omega)\Theta(\omega,T)$ (red curve) for the passive, unmodulated structure when the two sides are at the same temperature of $300$ K. It is seen that $\Phi_f(\omega)=\Phi_b(\omega)$ for all frequencies, as dictated by electromagnetic reciprocity. On the other hand, in the presence of modulation, $\Phi_f(\omega)\neq\Phi_b(\omega)$ due to the presence of the active region in the structure on the cold side, where power is either consumed or generated. This is seen in Fig. \ref{fig:fullsystem}(b) for a modulation of $\delta=1.5$ and $\Omega=2\pi\cdot 1.64$ THz, where $\Phi_f(\omega)\Theta(\omega,T_c)$ and $\Phi_b(\omega)\Theta(\omega,T_h)$ differ significantly in their spectral shape. Strikingly different from passive structures, $\Phi_b(\omega)$ can be negative at some frequencies in such modulated structures. This is because a current source in the hot emitter at frequency $\omega$ generates photons that cross the vacuum gap and generate sideband photons at $\omega+n\Omega$, which in turn experience partial reflection back into the vacuum gap, resulting in negative values of Poynting flux at the sideband frequencies.\\

By integrating the spectral heat flux in Fig. \ref{fig:fullsystem}(b), we obtain $\Delta P=+0.2033 \ \mathrm{W/m^2}$, indicating that heat flows against the temperature gradient. Further, we obtain a total power input of $64.8 \ \mathrm{mW/m^2}$ using Eq. \eqref{eq:wexact}. This gives us a thermodynamic COP of $(\Delta P/\dot{W})-1 = 2.1373$. Therefore, the structure shown in Fig. \ref{fig:matching}(a) indeed achieves photonic refrigeration after integration over all frequencies and wavevectors. In Fig. \ref{fig:fullsystem}(c), we plot the COP obtained from this system as a function of the modulation strength $\delta$ for a fixed $\Omega=2\pi\cdot 1.64$ THz. We see that the system begins to exhibit cooling for $\delta > 0.87$, saturating at a COP of around 2.3 for large modulation strengths. The performance of the full system is below the ideal limit of Fig. \ref{fig:intro} and the single-channel case of Fig. \ref{fig:matching} since the modes in the photonic crystal have varying frequency separations $\Omega$ and linewidths as the wavevectors (channels) are varied. In Fig. \ref{fig:fullsystem}(d), we plot the frequencies of modes 1 and 2 as a function of the angle of emission from the system. It is seen that the frequency separation between the modes varies from 1.64 THz to about 1.94 THz as the angle is varied. Due to this mismatch between the modulation frequency and the modal frequency separation, not all channels contribute equally to the cooling. Further improvements to the performance are possible by detailed optimization of the modal shapes involved in cooling, engineering the bands to be parallel over a larger angular range, considering non-planar geometries, as well as reducing the gap distance between the hot and cold sides to the near field regime. \\

We now comment on a few important aspects of our cooling approach. First, as compared to laser cooling, our cooling scheme enables a much higher COP. For a laser cooling system which up-converts a pump photon at energy $\hbar\omega$ to a luminescence photon at energy $\hbar\omega_f$, the COP can be expressed as \cite{seletskiy2016laser} $(\hbar\omega_f-\hbar\omega)/\hbar\omega \approx\ k_BT/\hbar\omega$, which inherently limits its COP to be on the order of 0.025 at room temperature for $\hbar\omega \simeq 1$ eV. By contrast, our approach can achieve a COP that is several orders of magnitude larger than laser cooling, reaching values close to the Carnot limit. In addition, our approach, being based on thermal emission, does not involve luminescence. Therefore, it can potentially overcome the stringent requirement on luminescence quantum efficiency in laser cooling and electroluminescent cooling. \\

To summarize, we introduce a new mechanism of active photonic refrigeration that is induced by temporal modulation of the refractive index in a thermal emission system. This mechanism has the potential to overcome the inherent performance limitations of laser cooling as well as the stringent luminescence requirements of electroluminescent cooling. In addition, our work also points to exciting new avenues for tailoring thermal emission. For example, the observed modulation-induced Rabi splitting and super-Planckian thermal emission provide a novel pathway for tuning thermal emission spectra for sensing applications. The combination of thermal photonics with temporal modulation opens up new avenues to tailor thermal emission for active cooling and energy harvesting applications. \\

This work was supported by Lockheed Martin, the U.S. Department of Energy Grant DE-FG-07ER46426, and the U.S. Department of Energy ``Photonics at
Thermodynamic Limits'' Energy Frontier Research Center under Grant DE-SC-0019140. S.B. acknowledges the support of a Stanford Graduate Fellowship. Valuable discussions with Avik Dutt and Professor David A.B. Miller are gratefully acknowledged.

\end{document}


\title{Photonic refrigeration from time-modulated thermal emission \\ Supplementary Material}
	
	\author{Siddharth Buddhiraju}
		\author{Wei Li}
	\author{Shanhui Fan}%
	\email{shanhui@stanford.edu}
	\affiliation{%
		Ginzton Laboratory, Department of Electrical Engineering, Stanford University, Stanford, CA
	}%
	\date{\today}
	\maketitle
	
\beginsupplement
\section{Coupled mode theory in the energy basis}
Consider a planar, layered structure along the $z$-direction with a dielectric function $\epsilon(z)$. A temporal modulation of the form $\delta(z)\cos(\Omega t)$ is applied to the structure, where $\Omega$ is the modulation frequency and $\delta(z)$ is the amplitude of the modulation that is non-zero only in the modulated layer(s). Then, Maxwell's equations in the structure take the form
\begin{align}
    \na\times\nabla\times E(z,t) +\frac{1}{c^2}\frac{\partial^2}{\partial t^2}\Big((\epsilon(z) + \delta(z)\cos(\Omega t))E(z,t)\Big) = 0, \label{eq:maxwell}
\end{align}
where $E(z,t)$ is the electric field in the structure and $c$ is the speed of light in vacuum.\\ 

Suppose the structure supports two modes with field profiles $E_{1,2}(z)e^{i\omega_{1,2}t}$ in the absence of modulation. When a weak modulation is applied, the electric field in the structure can be written as 
\begin{equation}
    E(z,t) = \tilde{a}_1(t)E_1(z)e^{i\omega_1 t} + \tilde{a}_2(t)E_2(z)e^{i\omega_2 t} \label{eq:fields}
\end{equation}
where $\tilde{a}_{1,2}(t)$ are slowly-varying amplitudes of the two modes. Inserting Eq. \eqref{eq:fields} into Eq. \eqref{eq:maxwell} and applying the rotating-wave approximation, we have from the $e^{i\omega_1 t}$ component in Eq. \eqref{eq:maxwell},
\begin{equation}
   -\frac{\omega_1^2}{2}\tilde{a}_2(t)\delta(z)E_2(z) + i\omega_1(\partial_t \tilde{a}_2)\delta(z)E_2(z) + 2i\omega_1(\partial_t\tilde{a}_1)\epsilon(z)E_1(z) = 0, \label{eq:derivation}
\end{equation}
where we use $\na\times\na\times E_{1,2}(z)=\omega_{1,2}^2E_{1,2}(z)/c^2$ for the unperturbed modes. Assuming that  $\partial_t \tilde{a}_{1,2}$ and $\delta(z)$ are both small, the second term in Eq. \eqref{eq:derivation} is a second order correction, whereas the other two terms are of the first order. Keeping only the first order terms, we get
\begin{equation}
    \epsilon(z)E_1(z)\partial_t \tilde{a}_1 = -\frac{i\omega_1}{4}\delta(z)E_2(z) \tilde{a}_2(t).
\end{equation}
Multiplying $E_1^*(z)$ on both sides and integrating over the structure, we obtain 
\begin{equation}
    -i\partial_t \tilde{a}_1 = V\sqrt{\frac{\omega_1}{\omega_2}}\tilde{a}_2(t),
\end{equation}
where we use the normalization $\int dz E_1^*(z)\epsilon(z)E_1(z)=1$ and define the coupling as $V = -\sqrt{\omega_1\omega_2}\int dz E_1^*(z)\delta(z)E_2(z)/4$. The same procedure applied at $\omega_2$ yields
\begin{equation}
    -i\partial_t \tilde{a}_2 = V^*\sqrt{\frac{\omega_2}{\omega_1}}\tilde{a}_1(t).
\end{equation}
Lastly, defining $a_{1,2}(t) = \tilde{a}_{1,2}(t)e^{i\omega_{1,2}t}$, we get
\begin{equation}
    -i\partial_t \begin{pmatrix} a_1 \\ a_2 \end{pmatrix} = \begin{pmatrix} \omega_1 & 0 \\ 0 & \omega_2 \end{pmatrix}\begin{pmatrix} a_1 \\ a_2 \end{pmatrix} + \begin{pmatrix} 0 & \sqrt{\frac{\omega_1}{\omega_2}}Ve^{-i\Omega t} \\  \sqrt{\frac{\omega_2}{\omega_1}}V^* e^{i\Omega t} & 0 \end{pmatrix}\begin{pmatrix} a_1 \\ a_2 \end{pmatrix} \label{eq:mod}
\end{equation}
The second term of Eq. \eqref{eq:mod} constitutes the time-modulation operator $M(t)$ used in Eq. (2) of the main text.   

\section{Thermal emission and absorption}
In this section, we derive quantities $P_{em}$ and $P_{abs}$ representing emission from and absorption by the cold side. The results of this section form the basis of Section IV, where the net cooling and the threshold condition shown in Eqs. (3)-(6) of the main text are derived. \\

The time evolution of the modes $a_{1,2}(t)$ in Eq. (1) of the main text is driven by noise sources on the cold side ($n_{1,2}^i$) as well as the hot side ($n_{1,2}^e$). On the other hand, the thermal emission from the cold side depends only on the noise sources $n_{1,2}^i$ on the cold side. Since the system in Eq. (1) is linear and the noise sources on the hot and cold sides are physically separated, a simple procedure to compute the thermal emission from the cold side involves removing the noise sources $n_{1,2}^e(t)$ corresponding to the fluctuations in the hot side. Then, the thermal emission from the cold side is given by \cite{otey2010thermal, zhu2013temporal} 
\begin{equation}
    P_{em} = 2\gamma_1^e\langle a_1^*(t)a_1(t)\rangle + 2\gamma_2^e\langle a_2^*(t)a_2(t)\rangle. \label{eq:em_form}
\end{equation}
The modes $a_{1,2}$ can be determined from Eq. (1) of the main text by a Fourier transform, resulting in $a_{1,2}(\omega)$ given by 
\begin{align}
    a_1(\omega) &= \frac{1}{Z_1(\omega)}\left(\sqrt{2\gamma^i_1}(\omega-\omega_1-i\gamma_2)n_1^i(\omega) + \sqrt{\frac{\omega_1}{\omega_2}} V\sqrt{2\gamma_2^i}n_2^i(\omega+\Omega) \right) \nonumber\\ 
a_2(\omega) &= \frac{1}{Z_2(\omega)}\left(\sqrt{\frac{\omega_2}{\omega_1}} V^*\sqrt{2\gamma_1^i}n_1^i(\omega-\Omega) +\sqrt{2\gamma^i_2}(\omega-\omega_2-i\gamma_1)n_2^i(\omega) \right), \label{eq:fourier_amplitudes}
\end{align}
where $\gamma_{1,2} = \gamma_{1,2}^i + \gamma_{1,2}^e$ and $Z_{1,2}(\omega) = (\omega-\omega_{1,2}-i\gamma_1)(\omega-\omega_{1,2}-i\gamma_2) - |V|^2$. 
Using Eq. \eqref{eq:fourier_amplitudes} with the relations $\langle n_{1,2}^{i*}(\omega)n_{1,2}^i(\omega')\rangle = \Theta(\omega,T_c)\delta(\omega-\omega')$, we obtain the emitted power averaged over a modulation period of $2\pi/\Omega$ as
\begin{align}
    P_{em} &= 4\gamma_1^e\gamma_1^i\int_{-\infty}^\infty d\omega \frac{(\omega-\omega_1)^2 + \gamma_2^2}{|Z_1(\omega)|^2}\Theta(\omega,T_c) + 4\gamma_2^e\gamma_2^i\int_{-\infty}^\infty d\omega \frac{(\omega-\omega_2)^2 + \gamma_1^2}{|Z_2(\omega)|^2}\Theta(\omega,T_c)  \nonumber \\
&+  4\gamma_1^e\gamma_2^i \int_{-\infty}^\infty d\omega \frac{\omega_1}{\omega_2}\frac{|V|^2}{|Z_1(\omega)|^2}\Theta(\omega+\Omega,T_c) + 4\gamma_2^e\gamma_1^i \int_{-\infty}^\infty d\omega \frac{\omega_2}{\omega_1}\frac{|V|^2}{|Z_2(\omega)|^2}\Theta(\omega-\Omega,T_c). \label{eq:emission}
\end{align}
The first two terms in this expression correspond to the individual thermal emissions of the two modes, being proportional to $\gamma_1^i\gamma_1^e$ and $\gamma_2^i\gamma_2^e$, respectively. The latter two terms arise purely from time modulation and are proportional to the strength of the modulation, $|V|^2$. These terms are `non-Planckian', since the photon number at frequency $\omega$ in these terms is $\Theta(\omega\pm\Omega,T)$, representing down- and up-conversion processes, respectively. \\

Similar to the procedure for obtaining Eqs. \eqref{eq:em_form}-\eqref{eq:emission} above, to calculate the absorption by the cold side, we remove the noise sources $n_{1,2}^i$ and place the external sources $n_{1,2}^e$ at temperature $T_h$. Solving Eq. (1) of the main text again, we get 
\begin{align}
    a_1(\omega) &= \frac{1}{Z_1(\omega)}\left(\sqrt{2\gamma^e_1}(\omega-\omega_1-i\gamma_2)n_1^e(\omega) + \sqrt{\frac{\omega_1}{\omega_2}} V\sqrt{2\gamma_2^e}n_2^e(\omega+\Omega) \right) \nonumber\\ 
a_2(\omega) &= \frac{1}{Z_2(\omega)}\left(\sqrt{\frac{\omega_2}{\omega_1}} V^*\sqrt{2\gamma_1^e}n_1^e(\omega-\Omega) +\sqrt{2\gamma^e_2}(\omega-\omega_2-i\gamma_1)n_2^e(\omega) \right). \label{eq:fourier_amplitudes_abs}
\end{align}
The power absorbed by the cold side is then given by 
\begin{equation}
    P_{abs}=2\gamma_1^i\langle a_1^*(t)a_1(t)\rangle + 2\gamma_2^i\langle a_2^*(t)a_2(t)\rangle.
\end{equation}
Using the expressions for $a_{1,2}(\omega)$ from Eq. \eqref{eq:fourier_amplitudes_abs} with the relations $\langle n_{1,2}^{e*}(\omega)n_{1,2}^e(\omega')\rangle = \Theta(\omega,T_h)\delta(\omega-\omega')$, we obtain
\begin{align}
    P_{abs} &= 4\gamma_1^e\gamma_1^i\int_{-\infty}^\infty d\omega \frac{(\omega-\omega_1)^2 + \gamma_2^2}{|Z_1(\omega)|^2}\Theta(\omega,T_h) + 4\gamma_2^e\gamma_2^i\int_{-\infty}^\infty d\omega \frac{(\omega-\omega_2)^2 + \gamma_1^2}{|Z_2(\omega)|^2}\Theta(\omega,T_h)  \nonumber \\
&+  4\gamma_2^e\gamma_1^i \int_{-\infty}^\infty d\omega \frac{\omega_1}{\omega_2}\frac{|V|^2}{|Z_1(\omega)|^2}\Theta(\omega+\Omega,T_h) + 4\gamma_1^e\gamma_2^i \int_{-\infty}^\infty d\omega \frac{\omega_2}{\omega_1}\frac{|V|^2}{|Z_2(\omega)|^2}\Theta(\omega-\Omega,T_h). \label{eq:absorption}
\end{align}
Note that the form of the last two terms for absorbed power in Eq. \eqref{eq:absorption} differ from the corresponding terms in emitted power in Eq. \eqref{eq:emission} in their pre-factors.

\section{Work in coupled mode theory}
To compute the work due to the modulation, we first note that in the absence of internal and external losses as well as modulation, the total energy in the modes $a_{1,2}$, i.e., $|a_1|^2 + |a_2|^2$, is conserved. Therefore, the contribution to the change in energy from the modulation alone can be determined by computing $d_t(|a_1|^2+|a_2|^2)$ and finding the terms arising from the non-Hermitian time-modulation operator $M(t)$. Computing $d_t(|a_1|^2+|a_2|^2)$ using Eq. (1) of the main text, we obtain the power input due to $M(t)$ to be
\begin{equation}
    \dot{W} =  \frac{2\Omega}{\sqrt{\omega_1\omega_2}}\Imag\left[\langle Ve^{-i\Omega t}a_1^*(t)a_2(t)\rangle\right]. \label{eq:work_exp}
\end{equation}
To determine the work done by the modulation on the photons emitted from the cold side, we follow the same procedure as the previous section and remove the external noise sources $n_{1,2}^e$. Then, inserting $a_{1,2}(\omega)$ from Eq. \eqref{eq:fourier_amplitudes} into Eq. \eqref{eq:work_exp}, we obtain 
\begin{equation}
    \dot{W}_c = |V|^2 \int d\omega \left(\frac{\Omega}{\omega_1}\frac{4\gamma_1^i\gamma_2}{|Z_1(\omega)|^2}-\frac{\Omega}{\omega_2}\frac{4\gamma_2^i\gamma_1}{|Z_2(\omega)|^2}  \right)\Theta(\omega,T_c). \label{eq:suppwork1}
\end{equation}
Similarly, to determine the work done by the modulation on the photons received from the hot side, we first remove the noise sources $n_{1,2}^i$ and use $a_{1,2}(\omega)$ from Eq. \eqref{eq:fourier_amplitudes_abs} in Eq. \eqref{eq:work_exp}: 
\begin{equation}
    \dot{W}_h = |V|^2 \int d\omega \left(\frac{\Omega}{\omega_1}\frac{4\gamma_1^e\gamma_2}{|Z_1(\omega)|^2}-\frac{\Omega}{\omega_2}\frac{4\gamma_2^e\gamma_1}{|Z_2(\omega)|^2}  \right)\Theta(\omega,T_h). \label{eq:suppwork2}
\end{equation}
\section{Derivation of net cooling and threshold condition}

\begin{figure}
    \centering
    \includegraphics[scale=0.5]{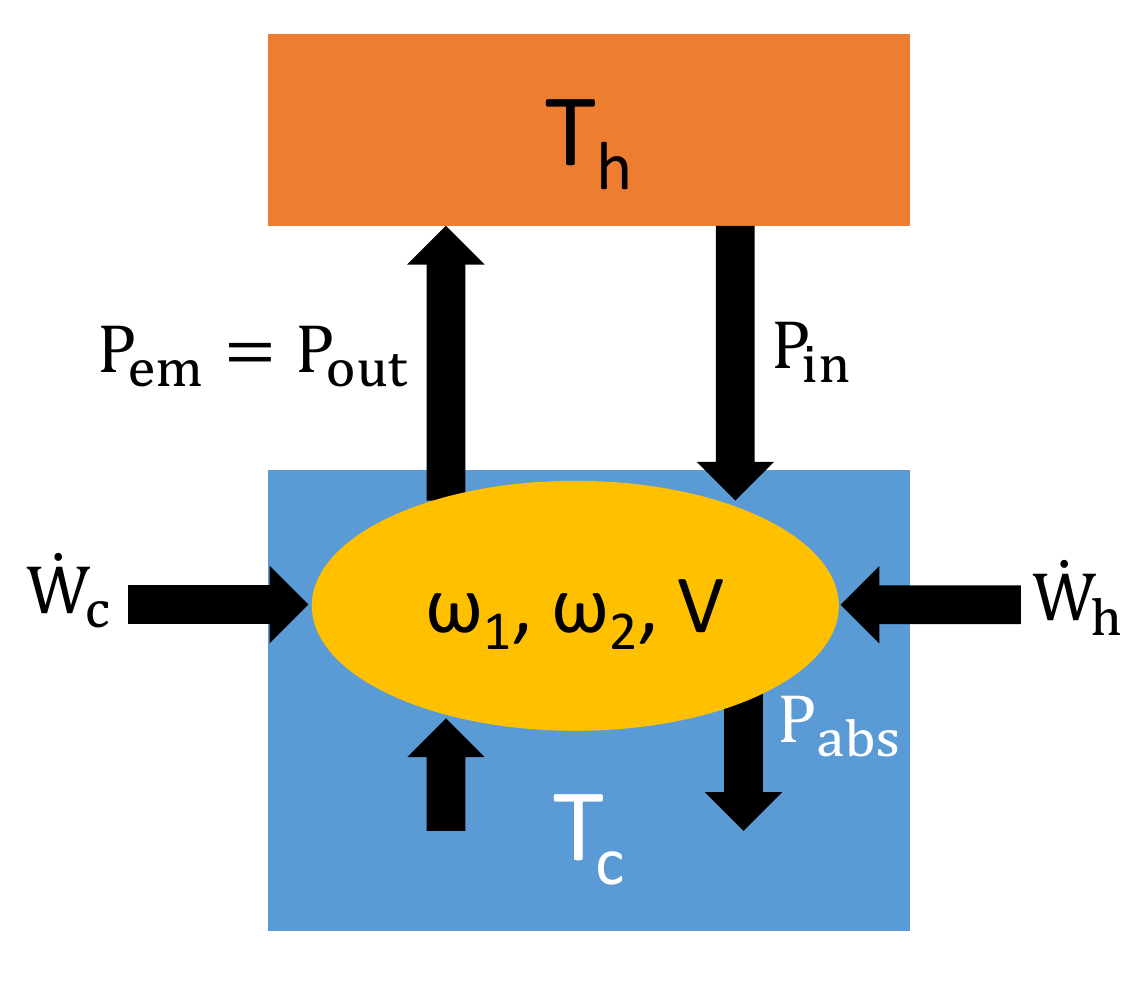}
    \caption{Schematic defining the quantities $P_{em}, P_{abs}, P_{out}, P_{in}$ and $\dot{W}_{c,h}$ for Sections IV and V. The emitted flux computed from Eq. \eqref{eq:emission} is $P_{em}$, which also represents the net outgoing power from the cold side. On the other hand, the absorbed flux computed from Eq. \eqref{eq:absorption} is $P_{abs}$, which is the power absorbed by the cold side. To compute the power incident on the cold side due to the hot side, $P_{in}$, we subtract the work done by the modulation, $\dot{W}_h$, on the incident photons.}
    \label{fig:SI_fig}
\end{figure}

Here, we derive Eqs. (3)-(6) of the main text. The results of this section are based on the emission, absorption and work calculations of Section II-III. Eqs. (3)-(6) of the main text correspond to $\gamma_1^i=\gamma_2^e=\gamma$ and $\gamma_1^e=\gamma_2^i=0$. Under these conditions, the power emitted by the cold side using Eq. \eqref{eq:emission} is
\begin{equation}
    P_{out}=P_{em} = 4\gamma^2 \int_{-\infty}^\infty d\omega \frac{\omega_2}{\omega_1}\frac{|V|^2}{|Z_2(\omega)|^2}\Theta(\omega-\Omega,T_c).
\end{equation}
This is identical to Eq. (3) of the main text. The power absorbed by the cold side can similarly be derived from Eq. \eqref{eq:absorption}. However, to obtain the power received from the hot side, we need to subtract the work done by the modulation on the photons incident from the hot side (see Fig. \ref{fig:SI_fig}):
\begin{equation}
    P_{in} = P_{abs}-\dot{W}_h = 4\gamma^2 \int_{-\infty}^\infty d\omega \frac{|V|^2}{|Z_1(\omega)|^2}\Theta(\omega+\Omega,T_h).
\end{equation}
This gives us Eq. (4) of the main text. Lastly, the total work input due to the modulation on the photons emitted from the cold side as well as the photons absorbed by the cold side is simply
\begin{equation}
    \dot{W}=\dot{W}_c+\dot{W}_h = 4\gamma^2|V|^2 \int \frac{d\omega}{2\pi} \left[\frac{\Omega}{\omega_1}\frac{\Theta(\omega,T_c)}{|Z_1(\omega)|^2} - \frac{\Omega}{\omega_2}\frac{\Theta(\omega,T_h)}{|Z_2(\omega)|^2}\right]. \label{eq:supp_work_total}
\end{equation}
where $\dot{W}_{c,h}$ are derived from Eqs. \eqref{eq:suppwork1}-\eqref{eq:suppwork2}. Eq. \eqref{eq:supp_work_total} is Eq. (5) of the main text. Using these expressions, we derive the threshold condition in the classical limit of $k_BT_{c,h} \gg \hbar\omega_{1,2}$. Defining $\int d\omega/|Z_{1,2}(\omega)|^2 = \alpha$, we have
\begin{align}
    P_{out} &= 4\gamma^2|V|^2\alpha \frac{\omega_2}{\omega_1} k_BT_c, \\ 
    P_{in} &= 4\gamma^2|V|^2\alpha k_B T_h, \\
    \dot{W} &= 4\gamma^2 |V|^2 \alpha\left(\frac{\Omega}{\omega_1} k_BT_c - \frac{\Omega}{\omega_2}k_BT_h\right).
\end{align}
Net cooling is achieved when $P_{out}-P_{in}-\dot{W}\ge 0$, which simplifies to
\begin{equation}
    T_c \ge \frac{\omega_1}{\omega_2} T_h.
\end{equation}
This is the same as the threshold condition in Eq. (6) of the main text.
\section{Carnot bound}
In this section, we provide a proof of the Carnot bound on cooling for a system described by the coupled-mode theory formalism in this paper. To calculate the coefficient of performance, we require that the system achieves net cooling. Recall that the power emitted by the system, $P_{em}$, is given by Eq. \eqref{eq:emission} and the power absorbed by the cold, $P_{abs}$, due to photons from the hot side is given by Eq. \eqref{eq:absorption}. Using the expression for work input, $\dot{W}_{c,h}$, from Eqs. \eqref{eq:suppwork1}-\eqref{eq:suppwork2}, we obtain the following thermodynamic quantities:
\begin{align}
    \dot{Q}_h &= P_{out}-P_{in} = P_{em}-(P_{abs}-\dot{W}_h), \\
    \dot{Q}_c &= (P_{em}-\dot{W}_c) - P_{abs} = P_{out}-P_{in}-\dot{W}, \\
    \dot{W} &= \dot{W}_c + \dot{W}_h,
\end{align}
where $\dot{Q}_c$ is the net cooling of the cold side, $\dot{Q}_h$ the net emission from the cold side and $\dot{W}$ the total work input due to the modulation. Note that $\dot{Q}_h = \dot{Q}_c + \dot{W}$, as required by the first law of thermodynamics. \\

Here, we prove that the quantity $\dot{Q}_c/\dot{W}$ is bounded by the Carnot limit of $T_c/(T_h-T_c)$. We assume the limit of $V\gg \gamma_{1,2}$ and that $\Theta(\omega,T)$ is approximately constant over the linewidth of the peak. This limit serves to provide a simplification to the integrals involved:
\begin{equation}
    \int_{-\infty}^\infty d\omega \frac{(\omega-\omega_{1,2})^2+\gamma_{2,1}^2}{|Z_{1,2}(\omega)|^2} \Theta(\omega,T)  = \int_{-\infty}^\infty d\omega \frac{|V|^2}{|Z_{1,2}(\omega)|^2} \Theta(\omega,T) \simeq \frac{1}{\gamma_1+\gamma_2}\Theta(\omega_{1,2},T).
\end{equation}
Further, we assume that $k_BT_{c,h}\gg \hbar\omega_{1,2}$ and hence  $\Theta(\omega_{1,2},T_{c,h})=k_BT_{c,h}$, allowing for simplicity of notation. However, the following proof also applies outside of this assumption due to the simple observation that $\Theta(\omega,T)/T$ is monotonically increasing in $T$. \\

Under these approximations, we get
\begin{align}
    P_{em} &= K\left[\gamma_1^e\gamma_1^i  + \gamma_2^e\gamma_2^i + \gamma_1^e\gamma_2^i\frac{\omega_1}{\omega_2} + \gamma_2^e\gamma_1^i\frac{\omega_2}{\omega_1}\right]T_c, \\ 
    P_{abs} &= K\left[\gamma_1^e\gamma_1^i  + \gamma_2^e\gamma_2^i + \gamma_2^e\gamma_1^i\frac{\omega_1}{\omega_2} + \gamma_1^e\gamma_2^i\frac{\omega_2}{\omega_1}\right]T_h, \\
    \dot{W}_c &= K\left[\frac{\Omega}{\omega_1}\gamma_1^i\gamma_2^e - \frac{\Omega}{\omega_2}\gamma_2^i\gamma_1^e \right]T_c + K\Omega\gamma_1^i\gamma_2^i\left[\frac{1}{\omega_1}-\frac{1}{\omega_2} \right]T_c \\
    \dot{W}_h &= K\left[\frac{\Omega}{\omega_1}\gamma_1^e\gamma_2^i - \frac{\Omega}{\omega_2}\gamma_2^e\gamma_1^i \right]T_h + K\Omega\gamma_1^e\gamma_2^e\left[\frac{1}{\omega_1}-\frac{1}{\omega_2} \right]T_h,
\end{align}
where $K = 4k_B/(\gamma_1+\gamma_2)$ is a constant. Notice that the second terms in the expressions for $\dot{W}_{c,h}$ are positive definite as $\omega_1<\omega_2$, and can be ignored since we wish to prove an upper bound on $\dot{Q}_c/\dot{W}$. Therefore,
\begin{align}
    \mathcal{C} &= \frac{\dot{Q}_c}{\dot{W}} \nonumber \\
    &\le \frac{(\gamma_1^i\gamma_1^e + \gamma_2^i\gamma_2^e)(T_c-T_h)+\gamma_1^i\gamma_2^e(T_c-\frac{\omega_1}{\omega_2}T_h) + \gamma_2^i\gamma_1^e(T_c-\frac{\omega_2}{\omega_1}T_h)}{\Omega\left(\gamma_1^i\gamma_2^e(\frac{T_c}{\omega_1}-\frac{T_h}{\omega_2}) + \gamma_2^i\gamma_1^e(\frac{T_h}{\omega_1}-\frac{T_c}{\omega_2}) \right)} \label{eq:step0} \\ 
    &\le \frac{\gamma_1^i\gamma_2^e(T_c-\frac{\omega_1}{\omega_2}T_h)}{\Omega\gamma_1^i\gamma_2^e(\frac{T_c}{\omega_1}-\frac{T_h}{\omega_2})} \label{eq:step1}\\
    &= \frac{\omega_1}{\Omega} \label{eq:step2}\\
    &\le \frac{T_c}{T_h-T_c}. \label{eq:step3}
\end{align}
Here, Eq. \eqref{eq:step1} follows from the assumption that the system achieves net cooling, implying that the numerator is net positive in Eq. \eqref{eq:step0}: the only term in the numerator that can be positive is the second term,  $\gamma_1^i\gamma_2^e(T_c-\frac{\omega_1}{\omega_2}T_h)$, when 
\begin{equation}
    T_c\ge \frac{\omega_1}{\omega_2}T_h. \label{eq:carnot}
\end{equation}
Consequently, the sum of all terms in the numerator of Eq. \eqref{eq:step0} must be smaller than the second term, leading to the numerator of inequality of Eq. \eqref{eq:step1}. Similarly, the second term in the denominator of Eq. \eqref{eq:step0} is positive definite and can be removed, giving the denominator of the inequality Eq. \eqref{eq:step1}. Lastly, the Carnot bound in Eq. \eqref{eq:step3} follows by imposing Eq. \eqref{eq:carnot} on the ratio $\omega_1/\Omega$.

\section{Work from Maxwell's equations}
In this section, we derive Eq. (10) of the main text. In the presence of periodic modulation with a modulation frequency $\Omega$, the electromagnetic field in the structure for a source excitation at frequency $\omega$ can in general be written as a Fourier series:
\begin{align}
    E(z,t) &= \sum_{n=-\infty}^\infty E_n(z) e^{i\omega_nt}, \nonumber \\
    H(z,t) &= \sum_{n=-\infty}^\infty H_n(z) e^{i\omega_nt}, \label{eq:field_forms}
\end{align}
where $\omega_n = \omega + n\Omega$, and $E_n(z)$ and $H_n(z)$ are the electric and magnetic fields corresponding to the $n$-th sideband, respectively. Written in terms of the coefficients $E_n$ and $H_n$, Maxwell's equations read
\begin{align}
    \na\times E_n &= -i\omega_n \mu_0 H_n \label{eq:floquet1}\\
    \na\times H_n &= +i\omega_n\sum_m\epsilon_{n-m}E_m, \label{eq:floquet2}
\end{align}
where $\epsilon_{n-m} = \frac{\Omega}{2\pi}\int_0^{2\pi/\Omega}\epsilon(t)e^{-i(n-m)\Omega t} dt$. The total time-averaged Poynting flux corresponding to the fields in Eq. \eqref{eq:field_forms} is $\sum_n \Real (E_n\times H_n^*)/2$. \\

To derive the work input due to modulation, notice that the modulated layer in Fig. 2 of the main text is lossless, leaving modulation to be the only source of breaking energy conservation. Therefore, any non-zero value of the Poynting flux integrated over a closed surface enclosing the modulated layer must correspond to work input due to the modulation. Using the Poynting theorem, we get for each sideband
\begin{equation}
    \oint \Real(E_n\times H_n^*)\cdot dS = \int \omega_n\Imag(E_n^*\cdot\sum_m\epsilon_{n-m}E_m) dV,
\end{equation}
where $\oint dS$ is integration over a surface enclosing the modulation layer and $\int dV$ over its volume. Defining $e = (...,E_{-1},E_0,E_1,...)^T$, $\hat{\epsilon}_{nm}=\epsilon_{n-m}$ and $\mathcal{W}_{nn}=\omega+n\Omega$, the total Poynting flux takes the form
\begin{align}
    \dot{W} \equiv \frac{1}{2}\sum_n \oint \Real(E_n\times H_n^*)\cdot dS = \frac{1}{2}\int dV \Imag\Tr\big[\mathcal{W}\hat{\epsilon}e e^\dag\big],
\end{align}
where the trace is over the sideband indices. Now, using $e(\rv)=\int d\rv'\mathcal{G}(\rv,\rv')j(\rv')$, where $\mathcal{G}(\rv,\rv')$ is the Green's function of the structure and $j(\rv')$ is a current source due to thermal fluctuations with the correlation $\langle j(\rv)j^*(\rv')\rangle = 4\omega\Theta(\omega,T)\Imag[\epsilon(\rv)]\delta_{\omega,\omega'}\delta(\rv-\rv')\delta(\omega-\omega')/\pi$, we obtain
\begin{equation}
    \dot{W} = \frac{2}{\pi}\omega\Theta(\omega,T)\int d\rv \int d\rv' \Imag\Tr\left[\mathcal{W\hat{\epsilon}G(\rv,\rv')}\delta_{\omega,\omega'}\Imag[\epsilon(\rv')]\mathcal{G^\dag(\rv,\rv')} \right]. \label{eq:pre_final}
\end{equation}
Here, $\delta_{\omega,\omega'}$ is a matrix that has a value 1 only on the diagonal element corresponding to the zeroth sideband and zero everywhere else. This form of the matrix ensures that thermal photons are generated only at $\omega$ because the lossy layers in the structure shown of Fig. 2 are unmodulated. The total work is then obtained by simply integrating Eq. \eqref{eq:pre_final} over all frequencies. 

\section{Coupled-mode theory parameters}
Here, we list the coupled-mode theory parameters for Fig. 2(b)-(g) of the main text. 
\begin{center}
    \begin{tabular}{c|c|c|c|c|c|c|c}
\hline
     Figure & $\omega_1$ (THz) & $\omega_2$ (THz) & $\gamma_1^i$ (GHz) & $\gamma_1^e$ (GHz) & $\gamma_2^i$ (GHz) & $\gamma_2^e$ (GHz) & $V$ (GHz)\\
     \hline\hline
     2(b)-(c) & 25.26 & 26.89 & 34.86 &  1.18 & 0.56 & 39.96 & 0\\
     2(d)-(e) & 25.26 & 26.93 & 33.78 &  1.20 & 0.52 & 42.84 & 11.58\\
     2(f)-(g) & 25.25 & 26.94 & 35.62 &  1.91 & 2.13 & 37.65 & 53.29\\
     \hline
\end{tabular}
\end{center}